\documentclass[a4paper]{article}
\usepackage[utf8]{inputenc}
\usepackage{xcolor}
\usepackage{xspace}
\usepackage{comment}
\usepackage{tipa}
\usepackage[T4,T1]{fontenc}
\usepackage{graphicx}
\graphicspath{ {./figs/} }
\usepackage{INTERSPEECH2022}
\usepackage{textcomp}
\usepackage{hyperref}
\usepackage[hang,flushmargin]{footmisc}
\usepackage{enumitem}

\AtBeginDocument{} %capitalize section
\AtBeginDocument{} %subsection->Section
\newcommand*{\ewe}{\'Ew\'e\xspace}
\newcommand*{\yoruba}{Yor\`ub\'a\xspace}
\newenvironment{tfour}{\fontencoding{T4}\selectfont}{}

 \title{BibleTTS: a large, high-fidelity, multilingual, and uniquely African speech corpus}
\name{Josh Meyer$^1$, David Ifeoluwa Adelani$^2$, Edresson Casanova$^{1,3}$, Alp \"Oktem$^{4,5}$, Daniel Whitenack$^{6}$ Julian Weber$^1$, Salomon Kabongo$^7$, Elizabeth Salesky$^8$, Iroro Orife$^9$, Colin Leong, Perez Ogayo$^{10}$, Chris Emezue$^{11}$, Jonathan Mukiibi $^{12}$, Salomey Osei, Apelete Agbolo$^{13}$,  Victor Akinode, \\Bernard Opoku$^{14}$, Samuel Olanrewaju, Jesujoba Alabi$^2$, Shamsuddeen Muhammad }
\address{
  $^\forall$ Masakhane NLP, Africa
  $^1$ Coqui, USA  
  $^{2}$ Saarland University, Germany
  $^3$ University of S\~ao Paulo, Brazil 
  $^4$ CLEAR Global, USA
  $^5$ Col·lectivaT, Spain
  $^6$ SIL International
  $^7$  Leibniz Universität Hannover, Germany
  $^8$ Johns Hopkins University, USA
  $^{9}$ Niger-Volta LTI, USA,
  $^{10}$ Carnegie Mellon University, USA
  $^{11}$ Technical University of Munich, Germany
  $^{12}$ Makerere University, Uganda\\
  $^{13}$ Ewegbe Akademi, Togo, 
  $^{14}$ Kwame Nkrumah University of Science and Technology, Ghana
  }
\email{josh@coqui.ai, didelani@lsv.uni-saarland.de, edresson@coqui.ai, alp.oktem@clearglobal.org, dan{\textunderscore}whitenack@sil.org}

\begin{document}

\maketitle
\begin{abstract}

%   BibleTTS is the largest, highest-quality open Text-to-Speech dataset for \textit{any}  language, and immediately unlocks the development of high-quality Text-to-Speech models for six languages spoken in Sub-Saharan Africa. 
  BibleTTS is a large, high-quality, open speech dataset for ten languages spoken in Sub-Saharan Africa. 
  The corpus contains up to 86 hours of aligned, studio quality 48kHz single speaker recordings per language, enabling the development of high-quality text-to-speech models. 
  The ten languages represented are: Akuapem Twi, Asante Twi, Chichewa, Ewe, Hausa, Kikuyu, Lingala, Luganda, Luo, and Yoruba. 
  This corpus is a derivative work of Bible recordings made and released by the Open.Bible project from Biblica. 
  We have aligned, cleaned, and filtered the original recordings, and additionally hand-checked a subset of the alignments for each language.
  We present results for text-to-speech models with Coqui TTS. 
  The data is released under a commercial-friendly CC-BY-SA license. 
\end{abstract}

% \section{Introduction} 
% There are multiple barriers to creating high-quality, neural Text-to-Speech (TTS). Access to data, compute, and engineering time are the key blockers to creating neural TTS models, and data is often the most significant. Most of the world's approximately 7,000 languages don't have any open datasets for speech technology, and even fewer have open, high-quality datasets for training TTS models. Great breakthroughs in machine learning are made possible by great datasets. MNIST, Imagenet, Common Voice, and LJSpeech have each heralded advancements in their respective fields, because the data is freely and openly available. In the field of TTS, it is notable that every new architecture is benchmarked on LJSpeech -- a dataset of American English. It is well known that the fields of natural language processing, speech-to-text and text-to-speech have a bias problem. The best open datasets are always in the English language, and typically in American English. We hope to advance the entire field of TTS and African TTS in particular with the release of the BibleTTS corpus. This data is higher recording quality than LJSpeech, and will allow researchers to make meaningful benchmarks against non-English languages.
% \textcolor{red}{Cite this: Google Paper that introduces a Yoruba dataset as justified:}% https://doi.org/10.21437/interspeech.2020-1096}

\section{Introduction} 

The majority of the world's approximately 7,000 languages~\cite{ethnologue} do not have open speech datasets, and even fewer have high-quality data with aligned text and speech, which can be used for training text-to-speech (TTS) models. 
The creation of benchmark datasets such as Librispeech~\cite{panayotov2015librispeech}, LibriTTS~\cite{zen2019libritts}, and LJSpeech~\cite{ljspeech17} enabled significant advances through community development on common resources, but these resources cover few languages, and most TTS systems evaluate on English only.

Speech synthesis systems have received significant attention in recent years due to the advances provided by deep learning. These advances enable TTS models to achieve improved naturalness with respect to human speech~\cite{tacotron2, valle2020flowtron, kim2021conditional}, and improved synthesized speech as driven adoption of virtual assistants~\cite{purington2017alexa, dempsey2017teardown}. However, neural models often require a non-trivial amount of data for training. This necessity leaves many language communities under-served in the development of speech technologies~\cite{casanova2022tts}, and it further results in researchers not evaluating models on diverse linguistic phenomena. 

In this work, we present the \textit{BibleTTS} corpus, a high-quality aligned speech corpus for ten African languages.  This data enables further research and resource creation for these languages and will allow researchers to create meaningful benchmarks against non-English languages.  

Creating high-quality aligned datasets typically requires tools not available for most languages, hindering the creation of datasets for lower-resourced languages.  Specifically, forced alignment of speech and text typically requires pre-trained acoustic models and grapheme-to-phoneme (G2P) models. This process can be challenging and error-prone without high-quality resources. We demonstrate that it is possible to force-align data without access to any pre-trained models (acoustic or G2P), and still produce quality output.

Additionally, recent corpora that significantly expand linguistic coverage for TTS datasets are often not freely available~\cite{black2019cmu,babel2011iarpa}, contain less single-speaker data~\cite{salesky2021mtedx}, and/or have lower-quality recordings. BibleTTS stands out in this regard as it is a large, high-fidelity corpus made of single-speaker recordings. The corpus is released under an open CC-BY-SA license. Corpus links and samples created with our TTS models can be accessed from the project website\footnote{\url{https://masakhane-io.github.io/bibleTTS/}}.

\section{Related Work} 

We focus on related work for African languages in the following section. 
Existing publicly available datasets are typically small. 
For \yoruba these include a 2.75 hour corpus~\cite{van2015lagos, Niekerk2012ToneRI} and a 4 hour multi-speaker dataset~\cite{gutkin2020developing}. TWB Gamayun kits \cite{gamayun} include a 6-hour single speaker high-quality Swahili speech corpus optimized for TTS training. 
Earlier \yoruba TTS efforts typically used bespoke private data~\cite{Dagba2016DesignOA, Afolabi2014DevelopmentOT, Akinwonmi2013APT, ajadi2007quantitative, Odejobi2004ACM}. 
For isiXhosa, Sesotho, Setswana and Afrikaans, multi-speaker corpora of approximately 2 hours each have been developed for TTS~\cite{van2017rapid, barnard2014nchlt}. 
The CMU Wilderness dataset \cite{black2019cmu} includes up to 20 hours of high-quality, single-speaker data for several African languages, but it is not publicly available and the alignments can contain noise.
TTS systems research for African languages has comprised development efforts in frameworks like Festival~\cite{Black1998FestivalSS} or MaryTTS~\cite{Schrder2011OpenSV} for \yoruba~\cite{iyanda2017development, aoga2016integration}, Ibibio~\cite{Ekpenyong2008TowardsAU},  Amharic~\cite{Mariam2004UnitSV}, Fon~\cite{Dagba2014ATT}, isiZulu~\cite{Louw2005AGI}, KiSwahili~\cite{Gakuru2005DevelopmentOA}. 
While many of these systems used concatenative synthesis, in large part because the available corpora were small, there have also been investigations into statistical parametric speech synthesis for Ibibio~\cite{Ekpenyong2014StatisticalPS}. 
Finally, there have been efforts in related tasks, such as grapheme-to-phoneme research for \yoruba~\cite{ynd2014DevelopmentOG}, intonation modeling~\cite{ajadi2007quantitative}, and numeral preprocessing~\cite{akinade2014computational}. %, to improve TTS for African languages. 

\section{Languages represented}

Table~\ref{tab:lang_statistics} shows the languages in the BibleTTS corpus, with their language families, the number
of speakers\cite{ethnologue} and the regions in Africa where they
are spoken. 
The corpus consists of ten languages from the three largest language families in Africa (Niger-Congo, Afro-Asiatic and Nilo-Saharan) and four regions of Africa. 
% Five of the languages (\ewe, Hausa, Akuapem Twi, Asante Twi, and \yoruba) are spoken in West Africa, three are spoken in East Africa (Kikuyu, Luganda, and Luo), one each in Central (Lingala) and South East (Chichewa) regions of Africa. 
All of these languages are tonal and are spoken primarily in sub-Saharan Africa. 

\subsection{Language Characteristics\protect\footnote{\texttt{https://omniglot.com/writing/{\textless}language-name{\textgreater}.htm}}}

\begin{comment}

\begin{table*}[t]
    \caption{Language and classification and statistics. All language classifications and number of speakers are from Ethnologue.}
    \label{tab:lang_statistics}
    \centering
    \resizebox{\textwidth}{!}{%
    \setlength\tabcolsep{16pt} %default 6pt
    \begin{tabular}[t]{llllr}
    \toprule
     & \bf  &  & \bf African & \bf No. of \\ 
    \bf Language & \bf ISO 639-3  &\bf Classification  & \bf  Region & \bf speakers \\ 
    \midrule
      \ewe        & ewe &  Niger-Congo / Kwa     & West    &  5.5M  \\
      Hausa       & hau &  Afro-Asiatic / Chadic & West    &  77M  \\
      Kikuyu      & kik &  Niger-Congo / Bantu   & East    &   8.2M \\
      Lingala     & lin &  Niger-Congo / Bantu   & Central &  40M \\
      Luganda     & lug &  Niger-Congo / Bantu   & East    &  11M \\
      Luo         & luo &  Nilo-Saharan / Luo–Acholi & East   &  5.3M\\
      Chichewa    & nya &  Niger-Congo / Bantu       &  South-East  & 14M\\
      Akuapem Twi & aka (Akuapem dialect) &   Niger-Congo / Akan &  West  & 626k \\
      Asante Twi  & aka (Asante dialect) &    Niger-Congo / Akan   &  West  & 3.8M \\
      \yoruba     & yor & Niger-Congo / Volta-Niger  &  West  & 46M \\
    \hline
\end{tabular}}%
\end{table*}
\end{comment}

\begin{table}[t]
    \caption{Language,  classification and statistics. All language classifications and numbers of speakers are from Ethnologue.}
    \vspace{-2mm}
    \label{tab:lang_statistics}
    \centering
    %\resizebox{\textwidth}{!}{%
    \scalebox{0.8}{
    \begin{tabular}[t]{lllr}
    \toprule
     & \bf  & \bf African & \bf No. of \\ 
    \bf Language  &\bf Classification  & \bf  Region & \bf speakers \\ 
    \midrule
      \ewe [ewe]    &    Niger-Congo / Kwa     & West    &  5.5M  \\
      Hausa [hau]    &  Afro-Asiatic / Chadic & West    &  77M  \\
      Kikuyu [kik]     &  Niger-Congo / Bantu   & East    &   8.2M \\
      Lingala [lin]     &  Niger-Congo / Bantu   & Central &  40M \\
      Luganda [lug]  &  Niger-Congo / Bantu   & East    &  11M \\
      Luo [luo]     &  Nilo-Saharan / Luo–Acholi & East   &  5.3M\\
      Chichewa [nya]  &  Niger-Congo / Bantu       &  South-East  & 14M\\
      Akuapem Twi [aka] &   Niger-Congo / Akan &  West  & 626k \\
      Asante Twi [aka]   &    Niger-Congo / Akan   &  West  & 3.8M \\
      \yoruba [yor]  & Niger-Congo / Volta-Niger  &  West  & 46M \\
    \hline
\end{tabular}}%
\vspace{-2mm}
\end{table}

\noindent \textbf{\ewe} [ewe] uses 35 Latin letters excluding (c, j, q), with 12 additional letters (\textrtaild, dz, \textepsilon, \textflorin, gb, \textgamma, kp, ny, \textipa{\ng}, \textopeno, ts, \textscriptv). Ewe has three tones, and they are marked in text. 

\noindent \textbf{Hausa} [hau] uses two different writing scripts: Ajami and Boko. The Boko script is the most widely used and is based on the Latin alphabet with 44 letters. The alphabet excludes letters (p, q, v and x) and uses 12 additional letters: \texthtb, \texthtd, \texthtk, \begin{tfour}\m{y}\end{tfour}, kw, {\texthtk}w, gw, ky, {\texthtk}y, gy, sh, ts. Hausa is tonal, but tones are not represented in text.

\noindent \textbf{Kikuyu} [kik] uses Latin script with 27 letters excluding (f, l, p, s, v, x, y, z), and including additional nine letters (\~{i}, \~{u}, mb, nd, nj, ng, ng`,ny, th). Kikuyu uses two tones (high and low) but they are not marked in text. 

\noindent \textbf{Lingala} [lin] uses the Latin script with 40 letters excluding (j, q, x) and including an additional 17 letters (\textepsilon, gb, kp, mb, mf, mp, mv, nd, ng, ngb, nk, ns, nt, ny, nz, \textopeno, ts). Lingala uses two tones (high and low), but they are not marked in text. 

\noindent \textbf{Luganda} [lug] uses 24 Latin letters excluding (h, q, x), and including additional two letters (\textipa{\ng}, ny). Luganda uses three tones, but they are not marked in text. 

\noindent \textbf{Luo} [luo] or Dholuo uses Latin script with 31 letters excluding the letters (c, q, x, v, z), and additional letters (ch, dh, mb, nd, ng’, ng, ny, nj, th, sh). Luo has four tones, but they are not marked in text.

\noindent \textbf{Chichewa} [nya] uses the Latin script with 31 letters excluding (q, x, y), and including additional eight letters (ch, kh, ng, \textipa{\ng}, ph, tch, th, \^{w}). Chichewa uses two tones (high and low) but they are not marked in text. 

\noindent \textbf{Akan} [aka] is a language with multiple dialects (including Fante, Bono, Asante, and Akuapem), and they are collectively known as Twi. In this study, we focus on Asante and Akuapem which are mutually intelligible and share the same alphabets (referred to herein as \texttt{aka-Asante} and \texttt{aka-Akuapem}). Twi uses 22 Latin letters excluding (c,j,q,v,x,z), and including two additional letters (\textepsilon, \textopeno).

\noindent \textbf{\yoruba}  [yor] uses 25 Latin letters without the letters (c, q, v, x and z) and with additional letters ({\d e}, gb, {\d s}, {\d o}). \yoruba is a tonal language with three tones: low, middle, and high. These tones are represented by the grave (e.g. ``\`{e} ''), optional macron (e.g. ``\={e}'') and acute (e.g. ``\'{e}'') accents respectively but the mid tone is usually ignored in writing.

\section{Corpus creation}

The BibleTTS corpus consists of high-quality audio released as 48kHz, 24-bit, mono-channel FLAC files. 
Recordings for each language are under professional quality, close-microphone conditions (i.e., without background noise or echo). 
BibleTTS is unique among open speech corpora for the volume of data per speaker and suitability for TTS.
% To the best of our knowledge, there exists no open speech corpus for TTS with so much high-quality single speaker data, for \textit{any} language. \red{unfortunately, there are (few) others, so I've changed this to 'rare'}
The corpus consists of ten languages which are under-represented in today's voice technology landscape, both in academia and in industry. We release {train/dev/test} splits for each language, where \texttt{dev} is the Book of Ezra, \texttt{test} is Colossians, and \texttt{train} is all other books.

\begin{figure}[t]
\centering
 \vspace{-2mm}
\caption{Distribution of the sample length per language. Samples longer than 30s  and with fewer than 10 characters were removed, and outlier segments were detected and discarded as described in \autoref{sec:outlier-detection}. 
Lingala is a slight outlier with the majority of segments between 10 and 20 seconds, while the other five languages have segments centered at 5-10s each.}
\label{fig:sample_lengths}
\includegraphics[width=\columnwidth]{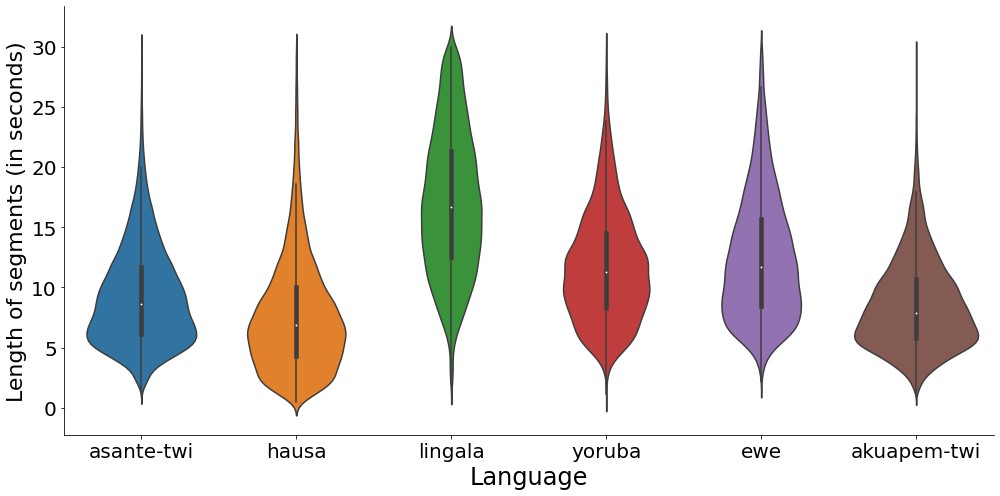}
\end{figure}

\subsection{Alignment}
\label{sec:alignment}

The BibleTTS corpus contains audio recordings and text transcripts (i.e. ``Open Contemporary Bible'' translations) which were released by Biblica via the Open.Bible project.\footnote{\url{https://open.bible/resources}} 
The original audio recordings were 48kHz, mono-channel WAV, typically one recording per chapter of the Bible. 
Each chapter was up to 30 minutes long, which is too long for most modeling tasks. 
Verses are a natural alternative, as the text already contains verse boundaries. 
Aligning at the verse level creates more manageable recording lengths of up to 30 seconds (see \autoref{fig:sample_lengths}) which are more likely to be consistent across languages than segmentation on voice activity detection or other alternatives. 

Potential challenges in alignment include additional content in either the speech or text, such as spoken titles and headings or text annotations, and the availability of pre-trained acoustic models and grapheme-to-phoneme mappings. 
We have employed various alignment techniques depending on the availability of verse timestamps and resources in each language, and evaluated a subset of the alignments with native speakers.

\subsubsection{Verse timestamps}

Three languages (\texttt{aka-Akuapem}, \texttt{aka-Asante}, and \texttt{lin}) were straightforwardly segmented using verse-level timestamps released by the Open.Bible project. The timestamps show the start time of every verse, as well as when the book and chapter titles were spoken. %, as below. 
With these timestamps, verses were isolated and saved as individual audio files using sox. These alignment scripts can be found on Github at \texttt{coqui-ai/open-bible-scripts}. \footnote{\url{https://github.com/coqui-ai/open-bible-scripts}}

% \begin{verbatim}
% Book Title 	00:00:02.00000000		
% Chapter Title	00:00:04.76000000		
% Heading 01	00:00:08.23331250		
% Verse 01	00:00:10.72827083		
% Verse 02	00:00:14.75897917	
% \end{verbatim}

\subsubsection{Forced alignment using pre-trained acoustic models}

Forced alignment is the process of extracting timestamps given an audio and a transcript pair, and requires either a pre-trained acoustic model or training one from scratch. For Hausa (\texttt{hau}), we opted to use the Montreal Forced Aligner (MFA) \cite{mfa} for which there is a pre-trained Hausa model \cite{globalphone}. The code is open-sourced on Github\footnote{\url{https://github.com/alpoktem/bible2speechDB}}. 
The process is as follows: 
\begin{enumerate}[leftmargin=*]
\itemsep-0.2em 
    \item Audio of each chapter of each book is downloaded together with their script in the form of an XML file, 
    \item XML script is parsed and converted into a plain normalized text file. Normalization entails: (a) adding the chapter title at the beginning of the script as "Sura <chapter-no>", 
    (b) converting numbers into written form using a dictionary prepared with Hausa linguists, and
    (c) adding a new line after every sentence ending punctuation mark (e.g. \texttt{.?!"}). 
    
    \item A grapheme-to-phoneme (G2P) dictionary is created from the word list extracted from the transcripts using the Hausa G2P model. 
    \item Alignment is performed for each chapter using the audio and normalized script with a beam length of 1000. 
    \item The time-aligned TextGrid file is processed in parallel with the sentence-segmented transcript to partition the chapter audio into sentence-level audio chunks with their transcriptions.
\end{enumerate}
\begin{comment}
\begin{enumerate}[label=(\alph*)]
\item adding the chapter title at the beginning of the script as "Sura <chapter-no>", 
\item converting numbers into their written forms using a dictionary prepared with Hausa linguists, and 
\item adding a new line after every sentence ending punctuation mark like period (.), question mark (?), exclamation mark (!) and quotation marks. 
\end{enumerate}
\end{comment}
% (1) Audio of each chapter of each book is downloaded together with their script in the form of an XML file, 
% (2) XML script is parsed and converted into a plain normalized text file. Normalization entails (a) adding the chapter title at the beginning of the script as "Sura <chapter-no>", (b) converting numbers into their written forms using a dictionary prepared with Hausa linguists, and (c) adding a new line after every sentence ending punctuation mark like period (.), question mark (?), exclamation mark (!) and quotation marks. 
% (3) A grapheme-to-phoneme (G2P) dictionary is created from the word list extracted from the transcripts using the Hausa G2P model. 
% (4) Alignment is performed for each chapter using the audio and normalized script with a beam length of 1000. 
% (5) The time-aligned TextGrid file is processed in parallel with the sentence-segmented transcript to partition the chapter audio into sentence-level audio chunks with their transcriptions. 
%The code to align Hausa is open-sourced on Github\footnote{\url{https://github.com/alpoktem/bible2speechDB}}.

\subsubsection{Forced alignment from scratch}

Two languages (\texttt{ewe} and \texttt{yor}) were aligned via forced alignment from scratch. 
Using only the found audio and transcripts (i.e., without a pre-trained acoustic model), an acoustic model was trained and the data aligned with the Montreal Forced Aligner. 
Graphemes were used as a proxy for phonemes in place of G2P data. 
The code used to generate alignments can be found in the \texttt{coqui-ai/open-bible-scripts} repository.\footnote{\url{https://github.com/coqui-ai/open-bible-scripts}}

After forced alignment, we used regular expressions to pull out whole verses which were aligned such that silence occurred both at the beginning and the end of a verse. 
Segmenting out audio at the verse-level instead of splitting on silence may allow downstream TTS models to capture higher-level prosody.

\subsection{Outlier detection}
\label{sec:outlier-detection}

Following the alignment stage, we detected and removed outliers using the \texttt{data-checker} toolkit together with human judgments. The relevant code is open-sourced on Github at \texttt{coqui-ai/data-checker.}\footnote{\url{https://github.com/coqui-ai/data-checker}} 
First, all segments longer than 30 seconds, or less than 10 characters in the aligned transcript, were removed.
Then, the removal of outliers was performed and fine-tuned for each language independently until the major offending samples\footnote{"Major offending samples" was not explicitly defined, but refers to samples labeled by a non-native speaker of these languages as containing obvious mismatches between transcripts and speech.} were no longer encountered, as described below. 

Every pair of \texttt{<audio,transcript>} was assigned an "outlier score", and the most extreme outliers were removed. First, the ratio of transcript length (characters) to audio length (seconds) was calculated for each sample. Then a Gaussian distribution was estimated for all samples in a given language. Lastly, the number of standard deviations from the mean was calculated for each sample. Outliers were excluded if they existed more than \texttt{N} standard deviations away from the mean, where \texttt{N} was fine-tuned per language with an iterative human-in-the-loop approach, until minimal offending samples were encountered. For most languages, it was sufficient to exclude samples more than 3 standard deviations from the mean (or .2\% of the data). 
However, \texttt{yor} notably required more outliers removed to attain a quality dataset. 
The resulting distribution of segment lengths per language is shown in \autoref{fig:sample_lengths}.

\subsection{Human evaluation of alignment quality}

We facilitated human evaluation of both the alignment and the output of the TTS models. In total, we collected labels from 15 annotators (three per language) for \texttt{ewe}, \texttt{hau}, \texttt{lin}, \texttt{aka-Asante}, and \texttt{aka-Akuapem} and an additional five annotators for \texttt{yor}. 
To judge the quality of \texttt{{\textless}audio,transcript\textgreater} pairs from our alignments, we randomly sampled 50 example pairings of aligned transcripts and the corresponding audio clips across the train, dev, and test sets. Annotators selected the one option that best described the quality of the alignment: 
\begin{enumerate}[leftmargin=*]
\itemsep0em 
    \item Audio contains EXTRA words not in the transcript
    \item Audio is MISSING words that are in the transcript
    \item Audio is MISSING words AND includes EXTRA words
    \item No missing or extra words
\end{enumerate}

In cases where the labels corresponding to various annotators disagreed, we took the majority vote label. In cases where the number of labels was spread evenly among different choices, we noted these as "conflicting." 
Results of human evaluation are shown in \autoref{tab:human_eval_alignment}. 

As discussed in Section~\ref{sec:alignment}, some languages (\texttt{aka-Asante}, \texttt{aka-Akuapem}, \texttt{lin}) were segmented using existing verse-level timestamp files. Interestingly, annotators labeled these languages as having a high percentage of samples where the audio contains additional words not present in the aligned text. Aligning from scratch (\texttt{ewe, yor}) produced a greater proportion of segments with exact matches between speech and text than using forced alignment with a pre-trained acoustic model (\texttt{hau}). However, it should be noted that significantly more data was removed due to outliers for \texttt{yor}, and less data overall was aligned with \texttt{ewe, yor} (see the statistics for unaligned vs. aligned hours in 
Table~\ref{tab:data_statistics}). 

\begin{table}[t]
    \caption{Corpus statistics. The corpus consists of data for ten languages, of which six have been aligned and formatted for immediate use to train TTS models.}
    \vspace{-2mm}
    \label{tab:data_statistics}
    \resizebox{\columnwidth}{!}{%
    \centering
    \begin{tabular}[t]{lrrrr}
    \toprule
                 & \bf Unaligned & \bf Unaligned & \bf Aligned & \bf Aligned\\ 
    \bf Language & \bf Hours     & \bf Samples   & \bf Hours   & \bf Samples \\
    \midrule
      \ewe        & 100.1 & 1,167  & 86.8 & 24,957 \\
      Hausa       & 103.2 & 1,189  & 86.6 & 40,603 \\
      Kikuyu      & 90.6 & 1,189   & -- & -- \\
      Lingala     & 151.7 & 1,189  & 71.6 & 15,117 \\
      Luganda     & 110.4 & 1,189  & -- & -- \\
      Luo         & 80.4 &  1,189  & -- & -- \\
      Chichewa    & 115.9  & 1,162 & -- & -- \\
      Akuapem Twi & 75.7 & 1,189  & 67.1 & 28,238  \\
      Asante Twi  & 82.6 & 1,189  & 74.9 & 29,021 \\
      \yoruba     & 93.6 & 1,189  & 33.3 & 10,228\\
    \hline
\end{tabular}}%
\end{table}

\section{TTS Models}
\label{sec:tts}
%\texttt{Edresson and David's section goes here XXX}

% old version 
% To show the quality of our dataset we propose to train the speech synthesis model VITS \cite{kim2021conditional} in the six languages with the highest number of hours. The chosen languages are Akuapem Twi, Asante Twi, Ewe, Hausa, Lingala, and Yoruba. 

%To accelerate training, in every experiment, we use transfer learning. We start from a model trained 1M steps on LJSpeech\cite{ito2017lj} provide in the Coqui TTS repository\footnote{\url{https://github.com/coqui-ai/TTS}} and continue the training for approximately 110K steps for each one of the languages.

%The models were trained using an NVIDIA A100 SXM4 80GB with a batch size of 100. For the TTS model training and for the discrimination of vocoder HiFi-GAN \cite{kong2020hifi} we use the AdamW optimizer \cite{loshchilov2017decoupled}  with betas 0.8 and 0.99, weight decay 0.01 and an initial learning rate of 0.0002 decaying exponentially by a gamma of 0.999875~\cite{paszke2017automatic}. 

To experimentally validate the quality of our dataset, we train the VITS end-to-end speech synthesis model \cite{kim2021conditional} with the sampling rate 22050 Hz in the six aligned languages. % with the highest number of hours. 
The chosen languages are Akuapem Twi, Asante Twi, \ewe, Hausa, Lingala, and \yoruba. We chose VITS for its state-of-the-art naturalness and also for its robust alignment mechanism \cite{kim2020glow}. The model takes characters as input and does not require a phonemizer.

To accelerate training we use transfer learning. We start from a model pre-trained on LJSpeech \cite{ito2017lj} for 1M steps, which is available via the Coqui TTS repository\footnote{\url{https://github.com/coqui-ai/TTS}}. We continue training for approximately 110K steps for each one of the languages. We use the AdamW optimizer \cite{loshchilov2017decoupled} with betas 0.8 and 0.99, weight decay 0.01, and an initial learning rate of 0.0002 decaying exponentially by a gamma of 0.999875~ \cite{paszke2017automatic}. The models were trained using an NVIDIA A100 SXM4 80GB with a batch size of 100. 
All models are released in the Coqui TTS toolkit\footnote{\url{https://github.com/coqui-ai/TTS}}.

\subsection{Results and Discussion}
\label{sec:results}

We evaluated the synthesized speech using subjective judgments, averaged across multiple speakers. 
% To evaluate the quality of the TTS models trained on the BibleTTS corpus, we randomly sampled 50 example transcripts from the test set, where the synthesized voices were created using the models trained as described in Section~\ref{sec:tts}. %this is
We randomly sampled 50 segments from the in-domain test set, as well as out-of-domain corpora to test the models' ability to generalize to non-Bible contexts. 
The out-of-domain sentences were obtained from the NEWS corpus\footnote{\url{http://github.com/masakhane-io/lacuna_pos_ner}} (except for Akuapem Twi).
Annotators rated the quality of synthesized speech in terms of naturalness of voice and appropriateness of pronunciation for the particular language or dialect. 
Annotators selected from a 5-point Likert rating for each sample: 1 (bad), 2 (poor), 3 (fair), 4 (good), and 5 (excellent). 
The mean opinion scores (MOS) are shown in \autoref{tab:eval_tts}. 
We additionally use mel cepstral distortion (MCD) \cite{kominek2008mcd}, an automatic edit distance metric, to assess quality for the in-domain segments where we have reference speech, with dynamic time warping (DTW) to align the segments. 
MCD largely follows MOS: languages with better human judgments (higher MOS) typically have better (lower) MCD scores, though MCD can be misleading, as in Lingala.

The MOS judgments seem related to the goodness of alignment evaluations. That is, the language with the best alignments (\texttt{ewe}) was also rated the best MOS for speech synthesized from the resulting model. Similarly, the language with the worst alignments (\texttt{hau}) resulted in the TTS model with the lowest out-of-domain MOS scores. To improve MOS, it may be necessary to either improve the alignments or apply more stringent outlier exclusion criteria, which should be possible, as the training data size remains significantly larger than many available TTS corpora for these languages or others. 

\begin{table}[t]
    \caption{Human evaluation of alignment.  Shown are percentages of <audio,transcript> samples with an exact match (EM), added words, missing words, or both.}
    \vspace{-3mm}
    \label{tab:human_eval_alignment}
    \resizebox{\columnwidth}{!}{%
    \centering
    \begin{tabular}[t]{lrrrrr}
    \toprule
                 & \bf   & \bf  & \bf  &   & \bf Conflict. \\ 
    \bf Language & \bf EM     & \bf Add.   & \bf Miss.   & \bf Both & \bf Labels \\
    \midrule
      \ewe        & 92\% & 2\% & 4\% & 0\% & 2\% \\
      Hausa       & 32\% & 68\% &  0\% & 0\% & 0\% \\
      Lingala     & 74\% & 12\% & 0\% & 0\% & 14\% \\
      Akuapem Twi & 88\% & 0\% & 8\% & 2\% & 2\%  \\
      Asante Twi  & 78\% & 2\% & 6\%  & 6\% & 8\% \\
      \yoruba     & 76\% & 24\% & 0\% & 0\% & 0\% \\
    \hline
\end{tabular}}%
\end{table}

\begin{table}[t]
    % \caption{Human evaluation for TTS presented as an average of Likert ratings over annotators and samples.}
    \caption{Evaluation of TTS model outputs using both human judgments (MOS) and an automatic metric (MCD). In-Domain texts are Bible verses, and Out-of-Domain is news.}
    \vspace{-2mm}
    \label{tab:eval_tts}
    \resizebox{\columnwidth}{!}{%
    \setlength{\tabcolsep}{6pt}
    \centering
    \begin{tabular}[t]{lccc}
    \toprule
                 & \bf MOS & \bf MOS & \bf MCD \\
    \bf Language & In-Domain     &  Out-of-domain  &  \\
    \midrule
      \ewe        & 4.34 & 3.87  & 5.8 \\ % robotic
      Hausa       & 3.42 & 2.34  & 7.6 \\ % a little robotic
      Lingala     & 3.31 & 2.40  & 5.6 \\ % a little robotic
      Akuapem Twi & 2.79 & ---  & 7.5 \\ % a little robotic
      Asante Twi  & 3.07 & 2.44  & 6.8 \\ % robotic 
      \yoruba     & 4.06 & 2.93  & 5.8 \\ % a little robotic
    \hline
\end{tabular}}%
\end{table}

\section{Conclusions}

The BibleTTS corpus is the first of its kind in many respects. The quality and volume of the data is extremely rare in open speech corpora -- these are professional, studio quality, 48kHz recordings,
% Furthermore, the sheer volume of recordings per speaker is unprecedented in any open speech corpus, to the best of our knowledge. 
with up to 86 hours of verse-aligned data per language, for 10 languages spoken in sub-Saharan Africa. 
The BibleTTS license is research and commercial friendly: CC-BY-SA. We hope that this corpus will enable advances in speech technology for African languages and also will unlock new techniques in TTS, which require more, higher-quality data. 

We described our approach to verse and sentence-level alignment of the original found data with a variety of different resources. We used human evaluation to assess the quality of the resulting alignments, and validate the resulting data by training high-quality speech synthesis models with Coqui TTS. 

There are two clear and immediate avenues for future work: (1) verse-level alignment of the remaining four languages (\texttt{kik}, \texttt{lug}, \texttt{luo}, and \texttt{nya}), and (2) improvement of the quality of existing alignments. Given the volume of data per language, it may well be the case that we can be more conservative with outlier removal, keeping only 20 or 30 hours of the best data, and obtain even better resulting TTS models. Nevertheless, we have shown that the data can already be used to produce high-quality TTS models (as with Ewe), on both in and out of domain text. We plan to update BibleTTS such that we have high-quality verse-level alignments for all ten languages.

\section{Acknowledgements}
We are very grateful to the volunteers, Richard J. Bonnie, Komlanvi D. Akoly, Komlanvi M. Klove, Ibrahim Haruna, Oluwabusayo O. Awoyomi, Emmanuel Anebi, Christian Kilapi, Pacifick Taba, who helped with human evaluation and the Masakhane community. 
%\red{Thank any annotators which were not able to fit as authors due to the 20 limit.}

% Note: there is only one page for references, need to make sure the bibliography fits on one page. Can abbreviate conferences (ie EEE International Conference on Acoustics, Speech and Signal Processing -> ICASSP) rather than remove references until absolutely needed
\newpage
\bibliographystyle{IEEEtran}

\bibliography{mybib}

\begin{thebibliography}{10}
\providecommand{\url}[1]{#1}
\csname url@samestyle\endcsname
\providecommand{\newblock}{\relax}
\providecommand{\bibinfo}[2]{#2}
\providecommand{\BIBentrySTDinterwordspacing}{\spaceskip=0pt\relax}
\providecommand{\BIBentryALTinterwordstretchfactor}{4}
\providecommand{\BIBentryALTinterwordspacing}{\spaceskip=\fontdimen2\font plus
\BIBentryALTinterwordstretchfactor\fontdimen3\font minus
  \fontdimen4\font\relax}
\providecommand{\BIBforeignlanguage}[2]{{%
\expandafter\ifx\csname l@#1\endcsname\relax
\typeout{** WARNING: IEEEtran.bst: No hyphenation pattern has been}%
\typeout{** loaded for the language `#1'. Using the pattern for}%
\typeout{** the default language instead.}%
\else
\language=\csname l@#1\endcsname
\fi
#2}}
\providecommand{\BIBdecl}{\relax}
\BIBdecl

\bibitem{ethnologue}
\BIBentryALTinterwordspacing
D.~M. Eberhard, G.~F. Simons, and C.~D.~F. (eds.), ``Ethnologue: Languages of
  the world. twenty-third edition.'' SIL International, 2022. [Online].
  Available: \url{http://www.ethnologue.com}
\BIBentrySTDinterwordspacing

\bibitem{panayotov2015librispeech}
V.~Panayotov, G.~Chen, D.~Povey, and S.~Khudanpur, ``Librispeech: an asr corpus
  based on public domain audio books,'' in \emph{ICASSP}.\hskip 1em plus 0.5em
  minus 0.4em\relax IEEE, 2015.

\bibitem{zen2019libritts}
H.~Zen, V.~Dang, R.~Clark, Y.~Zhang, R.~J. Weiss, Y.~Jia, Z.~Chen, and Y.~Wu,
  ``Libritts: A corpus derived from librispeech for text-to-speech,''
  \emph{Interspeech}, 2019.

\bibitem{ljspeech17}
K.~Ito and L.~Johnson, ``The lj speech dataset,''
  \url{https://keithito.com/LJ-Speech-Dataset/}, 2017.

\bibitem{tacotron2}
J.~Shen, R.~Pang, R.~J. Weiss, M.~Schuster, N.~Jaitly, Z.~Yang, Z.~Chen,
  Y.~Zhang, Y.~Wang, R.~Skerrv-Ryan \emph{et~al.}, ``Natural tts synthesis by
  conditioning wavenet on mel spectrogram predictions,'' in
  \emph{ICASSP}.\hskip 1em plus 0.5em minus 0.4em\relax IEEE, 2018.

\bibitem{valle2020flowtron}
R.~Valle, K.~Shih, R.~Prenger, and B.~Catanzaro, ``Flowtron: an autoregressive
  flow-based generative network for text-to-speech synthesis,'' \emph{arXiv
  preprint arXiv:2005.05957}, 2020.

\bibitem{kim2021conditional}
J.~Kim, J.~Kong, and J.~Son, ``Conditional variational autoencoder with
  adversarial learning for end-to-end text-to-speech,'' in \emph{International
  Conference on Machine Learning}.\hskip 1em plus 0.5em minus 0.4em\relax PMLR,
  2021.

\bibitem{purington2017alexa}
A.~Purington, J.~G. Taft, S.~Sannon, N.~N. Bazarova, and S.~H. Taylor, ``"
  alexa is my new bff" social roles, user satisfaction, and personification of
  the amazon echo,'' in \emph{Proceedings of the 2017 CHI Conference Extended
  Abstracts on Human Factors in Computing Systems}, 2017.

\bibitem{dempsey2017teardown}
P.~Dempsey, ``The teardown: Google home personal assistant,'' \emph{Engineering
  \& Technology}, 2017.

\bibitem{casanova2022tts}
E.~Casanova, A.~C. Junior, C.~Shulby, F.~S.~d. Oliveira, J.~P. Teixeira, M.~A.
  Ponti, and S.~Alu{\'\i}sio, ``Tts-portuguese corpus: a corpus for speech
  synthesis in brazilian portuguese,'' \emph{Language Resources and
  Evaluation}, 2022.

\bibitem{black2019cmu}
A.~W. Black, ``Cmu wilderness multilingual speech dataset,'' in
  \emph{ICASSP}.\hskip 1em plus 0.5em minus 0.4em\relax IEEE, 2019.

\bibitem{babel2011iarpa}
\BIBentryALTinterwordspacing
M.~Harper, ``The {IARPA} {B}abel multilingual speech database,'' 2011.
  [Online]. Available:
  \url{https://www.iarpa.gov/index.php/research-programs/babel}
\BIBentrySTDinterwordspacing

\bibitem{salesky2021mtedx}
E.~Salesky, M.~Wiesner, J.~Bremerman, R.~Cattoni, M.~Negri, M.~Turchi, D.~W.
  Oard, and M.~Post, ``Multilingual tedx corpus for speech recognition and
  translation,'' in \emph{Proceedings of Interspeech}, 2021.

\bibitem{van2015lagos}
D.~van Niekerk, E.~Barnard, O.~Giwa, and A.~Sosimi, ``Lagos-nwu yoruba speech
  corpus,'' 2015.

\bibitem{Niekerk2012ToneRI}
D.~R. van Niekerk and E.~Barnard, ``Tone realisation in a yor{\`u}b{\'a} speech
  recognition corpus,'' in \emph{SLTU}, 2012.

\bibitem{gutkin2020developing}
A.~Gutkin, I.~Demirsahin, O.~Kjartansson, C.~E. Rivera, and
  K.~T{\'u}b{\`o}s{\'u}n, ``Developing an open-source corpus of yoruba
  speech,'' 2020.

\bibitem{gamayun}
A.~Öktem, M.~A. Jaam, E.~DeLuca, and G.~Tang, ``Gamayun - language technology
  for humanitarian response,'' in \emph{2020 IEEE Global Humanitarian
  Technology Conference (GHTC)}, 2020.

\bibitem{Dagba2016DesignOA}
T.~K. Dagba, J.~O.~R. Aoga, and C.~C. Fanou, ``Design of a yoruba language
  speech corpus for the purposes of text-to-speech (tts) synthesis,'' in
  \emph{ACIIDS}, 2016.

\bibitem{Afolabi2014DevelopmentOT}
A.~A. Afolabi, E.~O. Omidiora, and O.~T. Arulogun, ``Development of text to
  speech system for yoruba language,'' 2014.

\bibitem{Akinwonmi2013APT}
A.~E. Akinwonmi and B.~K. Alese, ``A prosodic text-to-speech system for
  yor{\`u}b{\'a} language,'' \emph{ICITST}, 2013.

\bibitem{ajadi2007quantitative}
O.~O. {\`A}j{\`a}d{\'\i}, ``A quantitative model of yor{\`u}b{\'a} speech
  intonation using stem-ml,'' \emph{INFOCOMP Journal of Computer Science},
  2007.

\bibitem{Odejobi2004ACM}
O.~{\`A}. Od{\'e}job{\'\i}, A.~J. Beaumont, and S.~H.~S. Wong, ``A
  computational model of intonation for yor{\`u}b{\'a} text-to-speech
  synthesis: Design and analysis,'' in \emph{TSD}, 2004.

\bibitem{van2017rapid}
D.~van Niekerk, C.~van Heerden, M.~Davel, N.~Kleynhans, O.~Kjartansson,
  M.~Jansche, and L.~Ha, ``Rapid development of tts corpora for four south
  african languages,'' 2017.

\bibitem{barnard2014nchlt}
E.~Barnard, M.~H. Davel, C.~van Heerden, F.~De~Wet, and J.~Badenhorst, ``The
  nchlt speech corpus of the south african languages.''\hskip 1em plus 0.5em
  minus 0.4em\relax Workshop Spoken Language Technologies for Under-resourced
  Languages (SLTU), 2014.

\bibitem{Black1998FestivalSS}
A.~W. Black, P.~A. Taylor, and R.~Caley, ``Festival speech synthesis system,''
  1998.

\bibitem{Schrder2011OpenSV}
M.~Schr{\"o}der, M.~Charfuelan, S.~Pammi, and I.~Steiner, ``Open source voice
  creation toolkit for the mary tts platform,'' in \emph{INTERSPEECH}, 2011.

\bibitem{iyanda2017development}
A.~R. Iyanda and O.~D. Ninan, ``Development of a yor{\'u}b{\`a} textto-speech
  system using festival,'' \emph{Innovative Systems Design and Engineering
  (ISDE)}, vol.~8, no.~5, 2017.

\bibitem{aoga2016integration}
J.~O. Aoga, T.~K. Dagba, and C.~C. Fanou, ``Integration of yoruba language into
  marytts,'' \emph{International Journal of Speech Technology}, 2016.

\bibitem{Ekpenyong2008TowardsAU}
M.~E. Ekpenyong, E.-A. Urua, and D.~Gibbon, ``Towards an unrestricted domain
  tts system for african tone languages,'' \emph{International Journal of
  Speech Technology}, 2008.

\bibitem{Mariam2004UnitSV}
S.~H. Mariam, K.~Prahallad, A.~W. Black, R.~Kumar, and R.~Sangal, ``Unit
  selection voice for amharic using festvox,'' in \emph{SSW}, 2004.

\bibitem{Dagba2014ATT}
T.~K. Dagba and C.~Y. Boco, ``A text to speech system for fon language using
  multisyn algorithm,'' in \emph{KES}, 2014.

\bibitem{Louw2005AGI}
J.~A. Louw, M.~H. Davel, and E.~Barnard, ``A general-purpose isizulu speech
  synthesizer,'' \emph{South African Journal of African Languages}, 2005.

\bibitem{Gakuru2005DevelopmentOA}
M.~Gakuru, F.~K. Iraki, R.~C.~F. Tucker, K.~B. Shalonova, and K.~Ngugi,
  ``Development of a kiswahili text to speech system,'' in \emph{INTERSPEECH},
  2005.

\bibitem{Ekpenyong2014StatisticalPS}
M.~E. Ekpenyong, E.-A. Urua, O.~Watts, S.~King, and J.~Yamagishi, ``Statistical
  parametric speech synthesis for ibibio,'' \emph{Speech Communication}, 2014.

\bibitem{ynd2014DevelopmentOG}
A.~R. {\`I}y{\`a}nd{\'a}, O.~A. Odejobi, F.~A. Soyoye, and O.~O. Akinad{\'e},
  ``Development of grapheme-to-phoneme conversion system for yor{\`u}b{\'a}
  text-to-speech synthesis,'' 2014.

\bibitem{akinade2014computational}
O.~O. Akinad{\'e} and d.~A. \d{O}d\d{\'e}j\d{o}b{\'\i}, ``Computational
  modelling of yor{\`u}b{\'a} numerals in a number-to-text conversion system,''
  \emph{Journal of Language Modelling}, 2014.

\bibitem{mfa}
M.~McAuliffe, M.~Socolof, S.~Mihuc, M.~Wagner, and M.~Sonderegger, ``{M}ontreal
  {F}orced {A}ligner: Trainable text-speech alignment using {K}aldi,'' in
  \emph{Proc. Interspeech}, Stockholm, Sweden, 2017.

\bibitem{globalphone}
T.~Schultz, N.~T. Vu, and T.~Schlippe, ``Globalphone: A multilingual text amp;
  speech database in 20 languages,'' in \emph{2013 IEEE International
  Conference on Acoustics, Speech and Signal Processing}, 2013.

\bibitem{kim2020glow}
J.~Kim, S.~Kim, J.~Kong, and S.~Yoon, ``Glow-tts: A generative flow for
  text-to-speech via monotonic alignment search,'' \emph{Advances in Neural
  Information Processing Systems}, 2020.

\bibitem{ito2017lj}
K.~Ito \emph{et~al.}, ``The lj speech dataset,'' 2017.

\bibitem{loshchilov2017decoupled}
I.~Loshchilov and F.~Hutter, ``Decoupled weight decay regularization,'' 2017.

\bibitem{paszke2017automatic}
A.~Paszke, S.~Gross, F.~Massa, A.~Lerer, J.~Bradbury, G.~Chanan, T.~Killeen,
  Z.~Lin, N.~Gimelshein, L.~Antiga \emph{et~al.}, ``Pytorch: An imperative
  style, high-performance deep learning library,'' \emph{Advances in neural
  information processing systems}, 2019.

\bibitem{kominek2008mcd}
J.~Kominek, T.~Schultz, and A.~W. Black, ``Synthesizer voice quality of new
  languages calibrated with mean mel cepstral distortion,'' in \emph{Spoken
  Languages Technologies for Under-Resourced Languages}, 2008.

\end{thebibliography}

\end{document}